\documentclass[12pt,a4paper]{article}
\usepackage{amssymb}
\newcommand{\eq}[1]{(\ref{#1})}

\begin{document}
\title{Long-lived neutralino and ultra-high energy cosmic
rays.\footnote{Based on the talks given by S.T.\ at Quarks'2002 (the
12th International Seminar on High Energy Physics, Novgorod, Russia,
June 1-7, 2002) and at SUSY'02 (the 10th International Conference on
Supersymmetry and Unification of Fundamental Interactions, June 17-23,
2002, Hamburg, Germany).}  }

\author{D.S. Gorbunov and S.V. Troitsky \\
\small\em
Institute for Nuclear Research of the Russian Academy of
Sciences,\\
\small\em
60th October Anniversary Prospect 7a, 117312, Moscow, Russia
}
\date{}
\maketitle

\begin{abstract}
Secondary photons from decays of metastable neutralinos can contribute
to the ultra-high energy cosmic ray flux. The neutralino production rate
is too low in acceleration mechanisms to affect the cosmic ray
spectrum without emitting enormous energy in photons and
neutrinos. However, in top-down models with sources
not concentrated in galactic halos, neutralino decays change the
spectrum significantly. We estimate the parameters of a model in which
photons from neutralino decays are responsible for cosmic ray events
with energies above $10^{20}$~eV, and figure out distinctive
experimental signatures for this model.
\end{abstract}

{\bf 1.} Current experimental data on ultra-high energy (UHE) cosmic ray
(CR) spectrum are controversial. The results of the AGASA experiment \cite{AGASA}
indicate the absence of the GZK feature \cite{GZK}, a cutoff in the
spectrum at energies above a few times $10^{19}$~eV due to the attenuation of
high energy protons on cosmic background radiation. However,
recently published spectrum obtained by the HIRES experiment \cite{HIRES}
exhibits this cutoff. Other experiments had either low statistics or
insufficient precision to clearly support either the absence or the presence of
the GZK cutoff in the spectrum.

On the other hand, {\em all} experiments (including HIRES) have observed
cosmic rays with energies as high as $(1\div 3)\cdot 10^{20}$~eV .
Recently, impressive correlations have been found \cite{BL} 
between the arrival directions
of UHECRs registered by the AGASA and Yakutsk experiments\footnote{The
HIRES
stereo data has not been published, nor are we aware of any analysis
of the data.} with
the positions of gamma-ray loud BL Lac type objects. This implies that a significant
fraction of the extremely energetic particles may originate at cosmological
distances. Attenuation on the background photons then excludes protons and
photons as possible candidates for these particles.

One of the suggested ways to solve the problem is to consider other
particles which do not attenuate significantly on the photonic
background. The energy attenuation length of photons is of the same
order as that of protons. Neutrinos are obvious candidates, but
neutrino primaries are excluded by the atmospheric shower development
\cite{nonuprimaries}. However, UHE neutrinos can scatter off the relic
antineutrinos (and vice versa) via the $Z$-resonance. The sites of
these so-called $Z$-bursts serve as secondary sources of photons and
nucleons of somewhat smaller but still very high energy. If these
scatterings take place at a distance from the Earth less than the
nucleon's and photon's energy attenuation length, the $Z$ decay
products could contribute to the observed UHECR flux
\cite{Zburst}. The resonant energy of a neutrino with mass $m_\nu$ is
\begin{equation}
E_{\rm res}\approx {4~{\rm eV}\over m_\nu}\cdot 10^{21}~{\rm eV}.
\label{Eres}
\end{equation}
Other proposals for non-attenuating particles include rather exotic light
supersymmetric hadrons \cite{F} and light axion-like particles
\cite{sgoldst}.

In this talk, we propose an alternative to the $Z$-burst mechanism which
shares a number of its features but does not require the extraordinarily
high particle energies obtained from Eq.~\eq{Eres}.  Like the neutrino,
the neutralino can travel for cosmological distances unattenuated \cite{berezinsky}.
The lightest superpartners of the
Standard Model particles can, in some supersymmetric models, decay
either to lighter gravitino and non-supersymmetric particles (if $R$
parity is conserved), or to non-supersymmetric particles
alone with $R$ parity
violated. These decays can occur more frequently in our cosmological
neighbourghood if the lifetime of the particle in the laboratory
frame is of order but somewhat less
than the age of the Universe.
This gives rise to the super-GZK secondary particles in a
way analogous to the $Z$ burst mechanism. We note that
neutralino-induced atmospheric showers
would be very similar to those induced by neutrinos and hence can be
excluded on an equal footing to the neutrino events.

The dominant decay mode of the neutralino is photonic in the models we
consider here. This means that a signature of this mechanism is
the presence of photon-induced atmospheric showers. Current experimental
data restrict the fraction of photonic showers to be less than
$(28\div 48)\%$ at the energies $E\lesssim 10^{19.5}$~eV
\cite{photonshowers}. However, at higher energies the bounds are much
weaker, $(50\div 67)\%$ \cite{photonshowers}. We will see below that
the most probable implementation of our mechanism is to explain the
super-GZK events in the framework of a top-down mechanism while
relating the events below $10^{20}$~eV to protons accelerated in
active galaxies \cite{HIRES,berez1}.  

{\bf 2.} To estimate the required neutralino lifetime and flux, we
roughly approximate the decay rate of neutralinos as well as the rate
of energy loss of photons to the exponentials of the distances
travelled by particles. We denote the width of decay (neutralino $\to$
photon $+\dots$) measured in the laboratory frame as $\Gamma$; and the
mean energy attenuation length of a photon on the cosmic IR and radio
background as $l\sim 100$~Mpc. We suppose that the sources are
distributed in the Universe with the evolution index $m$ in the
comoving frame,
$$
dn(r)=n_0 4 \pi r^2 \left(1+z(r)\right)^m dr, ~ r<R,
$$
where $r$ is the distance from the Earth, $z(r)$ is the corresponding redshift.

The total UHE photon flux on the Earth, $n_\gamma$, can be expressed
via the total neutralino flux from all sources, $n_\chi$, as
$$
n_\gamma=n_\chi \,{\Gamma\over\Gamma-1/l}
{
\int_0^R\! dr\, r^2\, \left(1+{r\over R_0}\right)^{-2m-6} 
\left({\rm e}^{-r/l}-{\rm e}^{-\Gamma r}\right)
\over
\int_0^R\! dr\, r^2\, \left(1+{r\over R_0}\right)^{-2m-6} 
},
$$
where $R_0\approx4$~Gpc is the radius of the Universe, and we
calculated the fluxes in the laboratory frame.
For given $m$ and $R$, $n_\gamma/n_\chi$ has a broad maximum as a function
of $\Gamma$,
so the fine tuning of $\Gamma$ need
not be very strong. 
We present in Table 1 the neutralino lifetimes for three
sets of values of distribution parameters.
\begin{table}
\centerline{
\begin{tabular}{c|c|c|c}
\hline
$m$&$-2$ &0  &$+2$\\
\hline
$R$, Gpc& 1 &4&2  \\
\hline
maximal $n_\gamma/n_\chi$&0.086  &0.094  &0.068\\
\hline
$\tau$ at maximal  $n_\gamma/n_\chi$&0.5&6.8&2.9\\
\hline
\end{tabular}
}
\caption{ $R$ and $m$ are parameters of the source distribution,
$\tau$ is neutralino lifetime in the rest frame in units of $10^8~{\rm
s}\cdot\left({50~{\rm GeV} \over M}\right)$, $M$ is the neutralino mass. }
\end{table}
Note that the presence of a particle with lifetime
$\gtrsim 10^4$~s which decays to photons can affect Big Bang nucleosynthesis
due to subsequent photodisintegration of light nuclei \cite{nucleo}
unless the reheating temperature, and hence the particle abundance, are
low enough.

{\bf 3}. Let us turn now to specific supersymmetric models with metastable
neutralino. They consist of models with $R$ parity breaking where neutralino
LSP can decay to non-supersymmetric particles and models with gravitino
LSP with conserved $R$ parity (these include gauge-mediated supersymmetry breaking (GMSB)
\cite{GMSB} and certain no-scale supergravity models \cite{no-scale}). In what
follows, we will concentrate on GMSB scenario.

The lifetime of neutralino-NLSP in the restframe is
$$
\tau={16\pi^2\over \cos^2\theta_W}{F^2\over M^5},
$$
where $\theta_W$ is the weak mixing angle, $M$ is the neutralino mass,
and $F$ is the scale of dynamical supersymmetry breaking. The latter
is related to the gravitino mass, $m_{3/2}$, as
$$
F=\sqrt{3} M_* m_{3/2};
$$
$M_*=2.4\cdot 10^{18}$~GeV is the reduced Planck mass. We obtain
\begin{equation}
F=2.8\cdot 10^{19} ~ \mbox{GeV}^2 \left({\tau\over 10^8 {\rm s}}\right)^{1/2}
\left({M\over 50 ~{\rm GeV}}\right)^{5/2},
\label{F}
\end{equation}
\begin{equation}
m_{3/2}=6.5  ~ \mbox{GeV} \left({\tau\over 10^8 {\rm s}}\right)^{1/2}
\left({M\over 50 ~{\rm GeV}}\right)^{5/2}.
\label{gravitino}
\end{equation}
The gravitino is stable due to $R$ parity conservation and its mass
is constrained by the condition that relic gravitinos do not overclose the
Universe \cite{Moroi}. For $m_{3/2}$ in the GeV range and for reheating
temperature low enough to satisfy nucleosynthesis constraints on $\tau$,
the overclosure constraints are satisfied as well.

The values (\ref{F}), (\ref{gravitino}) are on the upper margins for usual
gauge mediation but can be natural in models of direct gauge mediation.
Indeed, let us consider probably the simplest complete model of GMSB
\cite{Agashe}. There, supersymmetry breaking is communicated directly from the
strongly interacting sector to the MSSM, and 
$$
M\approx {5\over 6\pi}\alpha_1(s){F\over s},
$$
where $s$ is the messenger scale, and $\alpha_1(s)$ is the $U(1)_Y$
coupling constant taken at the scale $s$. For $M\sim 50$~GeV and the values of
$F$ obtained above using Eq.(\ref{F}), this corresponds to $s\sim
10^{14}...10^{15}$~GeV depending on the required neutralino lifetime.
These values of $s$ are within the region allowed for the model of
Ref.\cite{Agashe} and low enough to suppress supergravity
contributions to soft masses with respect to GMSB contributions. 

{\bf 4.} 
The mechanisms responsible for creation of UHE particles can be divided
into three groups with distinctive observational signatures:

(1) acceleration in astrophysical sources -- arrival directions point back
to the sources, GZK cutoff is present in the spectrum assuming cosmological
distribution of the sources and protons or photons as UHE particles;
GZK cutoff is 
absent assuming non-attenuating UHE particles. This option seems to be
favoured by the data at energies below $10^{20}$~eV;

(2) the decay of metastable relic heavy particles or of short-lived heavy
particles originating in turn from the decay of metastable topological
defects which are distributed following the Cold dark matter (CDM) density: the sources are
concentrated in the halos of galaxies, and the dominant contribution to
the observed UHECR flux comes from the halo of the Milky Way (above GZK
energy, about 97\%  for nucleons and photons or $(15 \div 30)\%$ for non-attenuating
UHE partcles) \cite{DT}. Distribution of arrival directions exhibits
large-scale anisotropy due to the non-central position of the Sun in the Milky
Way \cite{DT}. GZK cutoff is absent in the spectrum;

(3) the decay of short-lived heavy particles originated in turn from
the decay
of metastable topological defects which do not follow the CDM
distribution
(an example of a
topological defect which does not follow the CDM density but is
distributed more or less homogeneously is provided by cosmic necklaces
\cite{berez-necklace}): 
 arrival directions of CRs are distributed uniformly (unless
there are only a few topological defects in the Universe), partial GZK cutoff
\cite{partialGZK} is present for protons or photons; it is absent for
non-attenuating particles.

We now consider different mechanisms of neutralino production and
check whether the mechanisms can produce the
required UHECR flux and not violate other observational constraints. We will
see that in acceleration mechanisms, option (1), required neutralino flux
can hardly be produced.

Indeed, the most probable mechanism of production of neutralinos in
astrophysical accelerators is in proton-proton collisions. For instance,
this could occur in the hot spots of active galaxies. All produced
supersymmetric particles
decay promptly to NLSP. To calculate the total
neutralino production cross-section, one thus has to sum over all
supersymmetric species. A collection of expressions for cross-sections can
be found in Ref.\cite{disserG}, and approximations for parton
distributions can be extrapolated from Ref.\cite{CTEQ}. At the energies
relevant to UHE production, the dominant SUSY production channel is gluon
fusion. We have checked that the partial cross-section
$\sigma_{\rm SUSY}/\sigma_{pp}\sim 10^{-8}$ at these energies, where we
extrapolated the total $pp$ cross-section from Ref.\cite{PDG}. If the UHE
protons do not escape from the source before collision with soft protons (this
is the case, for instance, in the hot spots of active galaxies), then the
total flux of UHE protons in all sources should exceed the observed UHECR
flux by a factor of about $10^9$:
$$
n_p\approx\left({\sigma_{SUSY}\over\sigma_{pp}}\right)^{-1} 
{n_\chi\over n_\gamma}n_\gamma
\sim 10^9 n_\gamma.
$$
The observed UHECR energy flux at energies $E\gtrsim 10^{20}$~eV is
$$
E^2J_{CR}(E)\approx 1 ~\mbox{eV cm$^{-2}$ s$^{-1}$ sr$^{-1}$}.
$$
The protonic flux of $10^9 E^2 J_{CR}(E)$ is excluded for the
following reasons. Firstly, the protons lose their energy by GZK
mechanism but do not disappear. Instead, they contribute to the CR
flux at lower energies \cite{WB}. The energy flux of protons at
sub-GZK energies is well measured and is only 10 times larger than the
flux at $10^{20}$~eV. Secondly, the dominant part of the energy flux in
$pp$ collisions is released roughly in equal amounts into photons and
neutrinos -- decay products of multiple $\pi$ mesons. The photons
lose their energy in electromagnetic cascades and contribute
\cite{emcascade} to the gamma ray background measured by EGRET
\cite{EGRET} which allows for $J_p/J_{CR}\lesssim 10^3$ and not
$10^9$. The constraints connected to protons and photons can in
principle be evaded by very high densities in the sources, so that the
protons do not leave the source at all\footnote{A similar problem
appears for the $Z$ burst mechanism if one assumes neutrino origin
from $pp$ collisions in astrophysical accelerators. In this case,
$J_p/J_{CR}\sim 10^4$ is required, and the on-site absorbtion can
help, though \cite{GTT} not in the most probable astrophysical
sources.}. However, neutrinos cannot be absorbed and overshoot the
current experimental limits (see Ref.~\cite{GTT} for a recent
compilation of data) by three orders of magnitude. The only
possibility to avoid neutrino production is to have enormous densities
in the source, $\sim 10^{19}$ protons$/$cm$^{-3}$. Then charged pions,
which carry about $2/3$ of the energy of the products of $pp$
collisions, would interact before their decay and lose energy in
pionic cascades so efficiently that neutrinos would be emitted only
with low energies. These neutrinos would contribute to a larger
atmospheric neutrino flux. These densities are hardly possible in
realistic astrophysical sources.

{\bf 5.} We conclude that in the context of the acceleration mechanism,
metastable neutralinos are irrelevant for UHECRs. On the other hand, in the
"top-down" mechanisms, supersymmetric particles (which promptly decay
to neutralino in our case) can carry about 40\% of the energy of the
original heavy particle \cite{berezinsky,neutralinoTD}. Photons from late
neutralino decays affect significantly the UHECR spectrum in the case of
homogeneously distributed sources (case (3)). 
The partial GZK cutoff inherent in these mechanisms is washed out
because neutralino decay probability is higher in our cosmological
neighbourghood. 
Currently, only a limited number of models of the type (3) are marginally
consistent with EGRET measurements \cite{EGRET} of gamma ray background
(see Ref.\cite{SiglTD} for examples of such models). With metastable
neutralinos, EGRET constraints are easily satisfied. The mechanism
discussed here has the following distinctive signatures in future
experiments: 

--- a neutralino which does not decay in the detector at future
    colliders but does not constitute the CDM;

--- the absence of positional correlations of CRs with specific astronomical
objects at energies $E>10^{20}$~eV;

--- global isotropy of arrival directions (including absence of galactic
anysotropy) at  $E>10^{20}$~eV;

--- high fraction of photons   at  $E>10^{20}$~eV.

In the case (2) of CDM-like distribution of the sources, the dominant part
of the UHECRs originate from decays of heavy particles within the Milky
Way. Unstable neutralinos can affect observable features of CRs in this
case only if they decay within the halo, that is their lifetime at rest is less
than $\sim 100$~s.

{\bf 6.} We are indebted to M.~Fairbairn, V.~Rubakov, D.~Semikoz, and
P.~Tinyakov for valuable discussions.  This work was supported in
part by the program SCOPES of the Swiss NSF, project No.~7SUPJ062239;
by CPG and SSLSS grant 00-15-96626; and by RFBR grant 02-02-17398. The
work of D.G. is supported in part by RFBR grant 01-02-16710 and INTAS
grant YSF~2001/2-142.  The work of S.T., as well as his participation
in SUSY-2002 conference, was supported in part by INTAS grant
YSF~2001/2-129. S.T.\ thanks Service de Physique Th\'eorique,
Universit\'e Libre de Bruxelles, for kind hospitality and support at
the final stages of the work.

\end{document}